\documentclass[usenatbib]{mn2e}
\input psfig.sty

\title[Structure and evolution of W UMa binaries with angular momentum loss]
{Structure and evolution of low-mass W UMa type systems -- II. with angular momentum loss}

\author[L. Li, Z. Han and F. Zhang]
{Lifang Li \thanks{E-mail:
gssephd@public.km.yn.cn or lifang\_li@hotmail.com}, Zhanwen Han and Fenghui Zhang\\
National Astronomical Observatories/Yunnan Observatory, Chinese Academy of
Sciences, P.O.Box 110,\\
Kunming, Yunnan Province 650011, P. R. China
}

\begin{document}

\date{Accepted yy mm dd. Received yy, mm, dd; in original form 2004 March 20}

\pagerange{\pageref{firstpage}--\pageref{lastpage}} \pubyear{2004}

\maketitle

\label{firstpage}

\begin{abstract}
In a preceding paper, using Eggleton's evolution code we have
discussed the structure and evolution of low-mass W UMa type
contact binaries without angular momentum loss (AML). The models
exhibit cyclic behavior about a state of marginal contact on a
thermal time-scale. Part of the time of each cycle is spent in
contact and part in a semi-detached state. According to
observations, W UMa systems suffer AML. We present the models of
low-mass contact binaries with AML due to gravitational wave
radiation (GR) or magnetic stellar wind (MSW) are presented. We
find that gravitational radiation cannot prevent the cyclic
evolution of W UMa systems, and the effect of gravitational
radiation on the cyclic behavior of contact binary evolution is
almost negligible. We also find that the most likely AML mechanism
for W UMa systems is magnetic braking, and that magnetic braking
effects can increase the period of the cyclic evolution, and
shorten the fraction of the time spent in the poor thermal contact
state exhibiting EB light curve. If W UMa stars do not undergo
cyclic evolution, and their angular momentum loss is caused
simultaneously by MSW of both components, we find that the value
of the parameter, $\lambda$, should be taken a larger value in
comparison with those derived from observations of single stars.
This indicates that the AML efficiency in W UMa systems may be
lowered in comparison with non-contact stars because of the less
mass contained in the convective envelopes of the components in W
UMa systems. If W UMa systems lose their angular momentum at a
constant rate. An angular momentum rate of $\frac{{\rm dln}J}{{\rm
d}t}\approx 1.6\times 10^{-9} {\rm yr^{-1}}$ can prevent the
cyclic behaviour of the model, and the model can keep in good
contact with an essentially constant depth of contact.

\end{abstract}


\section{Introduction}

In a preceding paper (Li et al. 2004, hereafter Paper I), the
structure and evolution of low-mass W UMa systems have been
discussed by us using the stellar evolution code developed by
\citet{egg71,egg72,egg73}, with updated physics \citep{pol95}. Our
models for low-mass W UMa systems were based on the conservative
assumption, namely that the total mass of the system and its
orbital angular momentum remain preserved. As a result, our models
exhibit cyclic behavior about a state of marginal contact with a
period of about $10^7$ yr. Part of the cycle is spent in contact
and part in the semi-detached state. In any cycle, a larger
temperature difference ($\Delta T_{\rm eff}>300$K) between the two
components occurs in a part of a cycle (about 30-35
per cent time of a cycle). So most previous investigators had
thought that this requires there to be as many short-period binaries
with EB light curves as with EW light curves, and that the TRO
models for W UMa type systems cannot explain the light curve paradox
according to the investigation of \citet{luc79}. \citet{ruc02}
finds 13 EWs and 5 EBs [and 14 ellipsoidal (ELL) variables, which
have too small an amplitude to be classified as EWs or EBs)]. It
is reasonable to identify the EWs as contact binaries, and the EBs
as semi-detached. The ratio of 5/13 is not much out of line with
TRO theory \citep{luc76, fla76,rob77} indicating that zero-age
contact systems are not thermally stable, but suffer cyclic,
thermally unstable mass transfer.

The TRO models for W UMa systems (Rahunen 1981, see also Paper I) has
shown that the predicted ratio of the timescales characterizing
the phases in which the oscillating systems will exhibit W
UMa-type (EW) and $\beta$ Lyrae-type (EB) light curves is
$\tau_{\rm EW}/\tau_{\rm EB}\leq 2$ (in the mass ratio range
$0.5\leq q\leq 0.7$). But the observations give  $\tau_{\rm
EW}/\tau_{\rm EB} \approx$ 5 to 6, suggesting that W UMa systems probably
suffer angular momentum loss. It is well known that loss of angular
momentum from the binaries consisting of two point masses via the emission
of gravitational wave radiation (GR) causes the separation between the
components to decrease. The possible importance of
gravitational radiation for short period binaries was first
pointed out by \citet{kra62}. In the evolution of W UMa systems
it has been discussed by \citet{fau71} and \citet{pac67}.
\citet{hua66} suggested that
magnetic torques could bring together the separate components of a
detached binary. This is Schatzman's (1962) and Mestel's (1968)
mechanism of magnetic braking. If the rapid rotation leads to
strong dynamos in the convective zones of the solar stars in W UMa
systems, effective magnetic braking may result and the angular
momentum loss by magnetic stellar wind (MSW) may control the contact binary evolution
\citep{vant79}. Meanwhile, there is increasing evidence of strong magnetic
activity in short-period binaries, including W UMa stars (spots, flares, strong
chromospheres and coronae, etc.).  This suggests indirectly that
W UMa systems suffer angular momentum loss via MSW. \citet{oka70} have considered the formation of W UMa
stars by magnetic braking in more detail. \citet{vant79,vant80}
has further developed this idea. The angular momentum loss in contact
binaries has also been treated by \citet{mos72}, \citet{web76}, \citet{rob77},
\citet{rah981} and \citet{vil82}.

The observational data and theory of contact binaries have been
reviewed in three extensive papers by \citet{moc81}, \citet{vil81} and
\citet{smi84}. Currently the prevailing opinion is that the AML
phenomenon is either the main property of or an ingredient in the
proper explanation of the existence and evolution of contact
binaries of late spectral types \citep{vant79,ruc82}. There is
ample evidence that the W UMa-type systems posses all the
chromospheric and coronal phenomena normally (i.e. in the solar
analogue) related to the existence of magnetic fields.
\citet{rob77} have found that a steady loss of angular momentum
drives the system towards smaller mass ratios. \citet{rah981} has
found that it is possible to keep zero-age models in good thermal
contact if the orbital angular momentum of the binary is allowed
to decrease and the timescale of the required angular momentum loss
is about $5\times 10^8$ yr.

In this paper, we restrict our attention to the general nature of the
effects of angular momentum loss (i.e. $\dot J<0)$ and to the influence of the
matter accreted by the secondary on its evolution in a semi-detached state
and present models of low-mass W UMa systems with angular momentum loss
via  gravitational radiation or magnetic braking. We find that the
influence of gravitational radiation on the cyclic evolution of
contact binaries is almost negligible because the
gravitational radiation life-time is long compared with the thermal timescale
of the primary of our models. Systemic angular momentum loss via gravitational
radiation only accelerates the evolution towards more extreme mass ratios,
rather than leading to orbital collapse. The most likely angular momentum loss
mechanism for W UMa systems is magnetic braking. It can increase the period
of the cyclic evolution and shorten the fraction of the time spent in the
state exhibiting EB light curves, and a suitable
angular momentum loss rate can prevent the cyclic behaviour of the model.
Meanwhile, if we consider an energy source at secondary's atmosphere provided
by the accreting matter from the primary in semi-detached
evolution, and find that the expansion in radius of the secondary is hastened
by the accreting matter which should have a higher entropy than the original
star, it can shorten the time spent in the state exhibiting EB light curve.
We compare the convective envelopes of a semi-detached model with
those of a contact one which has the same total mass and mass
ratio as the semi-detached one.
We find that the convective envelope of the secondary in the contact system
is much thinner than that in the semi-detached one and the convective envelope
of the primary in the contact system is slightly thicker than that in
semi-detached one. But the mass contained in the convective envelopes
of two components of contact system, which is related to the dynamo action,
is less than that contained in the envelopes of the semi-detached one.
This is caused by the energy transfer between the two components in contact
systems. As a result, W UMa systems show the lower activity in comparison
with the non-contact systems.

\section{Contact condition and Luminosity transfer}
\subsection{Contact condition}
In Paper I, based on Roche geometry, we have given a radius grid of contact
binaries. The radius grid gives the relative radii of both components of
contact binaries with different mass ratios (0.02, 0.04,...,1.0) and
different depth of contact (0.0, 0.025, 0.05,...,1.0) and it has been
used as the surface condition of contact binaries to ensure the surfaces of
both components lie on the same equipotential by interpolation. Here we use
the same contact condition as Paper I. If
the mass is transferred from star 2 to star 1, the rate of mass transfer
is expressed as,
\begin{equation}
\frac{{\rm d}m_2}{{\rm d}t}=-C {\rm Max}[0,({\rm ln}\frac{R_2}{R_{\rm 20}})^3]
\end{equation}
in which
\begin{equation}
R_{20}=\Re_{2} \root 3 \of {\frac{\varrho_{1}R_{1}^3+\varrho_{2}R_{2}^3}
{\varrho_{1}R_{1}^3+\varrho_{2}\Re_{2}^3}}.
\end{equation}
The explanations of Eqs. (1) and (2) are as in Paper I. If mass is transferred from star 1 to star 2, the rate of mass transfer
can be derived by the same way.

\subsection{Luminosity transfer}
\subsubsection{Luminosity transfer due to mass transfer}

In the semi-detached phase, the mass above the critical lobe, together with the
energy (including gravitational energy, heat energy, and
radiative energy) in the transferred mass, is transferred to the secondary.
The effect of heating the photosphere of a low-mass ZAMS stars (with
a deep convective envelope or fully convective), by the kinetic energy
of the infalling matter, is very important for their response \citep{pri85}.
Since the radiative energy is so small that it can be neglected relative to
gravitational energy and heat energy,
the accretion luminosity due to accreting mass is

\begin{equation}
L_{\rm acc}= \beta (\psi_{1}-\psi_{2}+s_{1}-s_{2})\dot{M_{2}}
\end{equation}
where $\psi_{i}$ and $s_{i}$ are, respectively, the gravitational potential
and specific entropy at the surface of the component $i$, $\dot{M_{2}}$ is the
rate of mass accretion of the secondary and $\beta$ is a coefficient which
describes the efficiency of heating of the photosphere by the shock wave region.
\citet{kah02a} took the parameter $\beta$ to be 1.0, i.e. he thought that
the bulk kinetic energy gained by falling through the potential is completely
transformed into thermal energy and it, together with the thermal energy carried
by infalling matter, is regarded as an energy source in the outermost layer
of the gainer without energy loss. However, some of this energy may be
dissipated dynamically in a shock, but the dissipated ratio of this energy
is unknown, so we used a value of $\beta= 0.5$ \citep{sar89}. This suggests
that 50 per cent of the transferred energy is lost from the system. Since
the accretion luminosity is caused by energy sources in the gainer's
outermost layers, we modify the surface boundary condition of the secondary
to model the influence of the accretion luminosity on the evolution of the
secondary, i.e. the surface temperature of the secondary may be approximated
by the formula, $L_{\rm in}+L_{\rm acc}=4\pi R^2 \sigma T_{\rm eff}^4$,
where $L_{\rm in}$ is the luminosity
coming to the photosphere from the stellar interior, $R$ is the radius
of the secondary, and $\sigma$ is the Stefan-Boltzmann constant.

\subsubsection{Luminosity transfer due to circulation currents in the common envelope}

\citet{str48} first recognized the unusual mass-luminosity
relationship of the secondary components of W Ursae Majoris
systems, i.e. the two components have identical temperatures and
hence luminosities scaling as $L\propto R^2$ ($\propto M$ for Roche
geometry) in spite of differing masses. This Suggests that it might be causally
related to a
possible common envelope. \citet{osa65} noted that von Zeipel's
theorem would require the observed approximate constancy of
radiative flux over the surface of a system with a radiative
common envelope in order to maintain hydrostatic equilibrium. The
fact that most W UMa systems appear to have convective envelopes,
however, led \citet{luc68} to propose that the anomalous mass-luminosity
relationship possessed by W UMa binaries can be explained if energy
transfer is assumed to take place between the two components of the systems
well below the atmosphere.

A luminosity transfer by circulation in the common envelope is applied
in the outermost layers which are close to the surface of each component. The amount of the luminosity
transfer by circulation must satisfy the requirement of W UMa systems'
observations
\citep{rob77}, it can be written as

\begin{equation}
 \frac{L_{1}-\Delta L_{0}}{m_{1}}=\frac{L_{2}+\Delta L_{0}}{m_{2}},
\end{equation}
where $L_{1,2}$ are the core luminosities (including nuclear and thermal luminosities) of both components, and $m_{1,2}$ the masses of the components. Since transfer is not fully efficient at all phases, an arbitrary factor
$f$ is introduced, which varies through the cycle and goes to zero with
the depth of contact. Thus we take:
\begin{equation}
\Delta L = f\cdot\Delta L_{0},\ \ \  (0\leq f\leq 1).
\end{equation}
where $f$ is the efficient factor of energy transfer. We take
\begin{equation}
f={\rm Min}[1,\alpha(d^{2}-1)]
\end{equation}
in which
\begin{equation}
d={\rm Max}[1,{\rm Min}(\frac{r_{1}}{R_{\rm crit1}},\frac{r_{2}}{R_{\rm crit2}})]
\end{equation}
where $r_{1,2}$ are the radii of both stars, $R_{\rm crit1,2}$ the
Roche critical radii of both stars. The parameter $\alpha$ is
expected to be moderately large, so that heat transfer becomes
fully efficient for stellar radii exceeding the Roche radii by
some standard small amount. In this work, we take $\alpha=15$.

The luminosity transferred by circulation currents from the
primary to the secondary adopted by \citet{kah02a} is
\begin{equation}
\Delta L = \int_0^{M_{2}} \sigma_{\rm ex,2} {\rm d}m_{2}
= -\int_0^{M_{1}}\sigma_{\rm ex,1}{\rm d}m_{1},
\end{equation}
where $\sigma_{\rm ex,i}$ is the source (when positive) or sink
(when negative) of energy per unit of mass caused by interaction
of the components. Although the luminosity increment is applied in the adiabatic
portion of the envelope of each star by most of the previous
investigators, according to the energy transfer model
\citep{rob80} there is no essential distinction between convective
and radiative envelopes. Meanwhile, Li et al. (Paper I) argued that
the energy transfer may take place in the outermost layers for
low-mass W-subtype systems and the transfer takes place
in the base of the common envelope for A-subtype systems. It follows
that convection is by no means essential to heat transport in
contact envelopes, although it may well have an important
influence. Therefore, we assume that the luminosity transfer takes
place in the outermost layers of the common envelope for low-mass
W UMa systems, and the luminosity increment is applied in the 10
meshpoints which are close to the surfaces of both components as
Paper I.

\section{Gravitational radiation and magnetic braking via MSW}

It is well known that loss of angular momentum from the binary
systems consisting of two point masses due to the emission of
gravitational radiation causes the separation between the
components to decrease \citep{lan62}. \citet{fau71} and \citet{taa80}
showed that gravitational radiation, by removing orbital angular
momentum, could stimulate mass transfer from the main sequence red
component at a rate which causes the binary to evolve more rapidly
than nuclear evolution would dictate. \citet{fau76} and \citet{cha77} have
both found very significant effects of $\dot{J}$ [via
gravitational radiation (GR) in their computations] on the
evolution of low-mass close binary systems. \citet{cha78} has
considered the effects of loss of orbital angular momentum via
gravitational radiation on the period changes and the implications
for mass-transfer rate determinations.

We adopt the standard Einstein theory which predicts that the
rates of change of semi-major axis and orbital angular momentum
are, respectively, \citep{lan62},
\begin{equation}
\frac{{\rm d}A}{{\rm d}t}=-\frac{64G^3}{5c^{5}A^3}M_1M_2(M_1+M_2)
\end{equation}
and
\begin{equation}
\frac{{\rm dln}J}{{\rm d}t}=-\frac{32G^3}{5c^{5}}\frac{M_1M_2(M_1+M_2)}{A^4},
\end{equation}
where $J$ is the orbital angular momentum, $A$ is the
semimajor axis of the system, $M_{\rm 1,2}$ are the
masses of both components, $G$ is the
gravitational constant, $c$ is the light velocity,
and $t$ is the time. The rate of angular momentum
loss associated with gravitational radiation is also given by \citep{web75},
\begin{equation}
\frac{{\rm dln}J}{{\rm d}t}=-8.1\
10^{-10}\frac{M_1M_2(M_1+M_2)}{A^4}\ {\rm yr^{-1}}
\end{equation}
where $J$ is the orbital angular momentum in ${\rm erg\cdot s}$, $M_{\rm 1,2}$ are the masses of both components in units of solar
masses $M_{\rm \odot}$, $A$ is the semimajor of the system in unit of
solar radii $R_{\rm \odot}$, $t$ is the time in years.

Magnetic braking of the rotation of the cool stars in the binaries is an attractive
mechanism. The idea that in stars with convective envelopes a dynamo is operating, which has efficiency that decreases with decreasing rotation rate, is consistent with observational data which indicate that the chromospheric (i.e. Ca II
H and K) and coronal emissions are strongest in rapidly rotating stars \citep{zwa81}. The magnetic braking is presumably due to the stellar wind which streams
out along the coronal magnetic field lines, which are rooted in the convection
zone \citep{sch62,sch65}. The loss of angular momentum per unit mass in the
stellar wind is very large since the outflowing matter is forced by magnetic
field to corotate with the star out to large distances.
The magnetic stellar wind (MSW) as a possible driving force for the evolution
of low-mass binaries was proposed first by \citet{hua66}. Observational
support for the action of such a wind in low-mass spectroscopic binaries has
been provided by \citet{kra78}. Theoretical studies of the effect of MSW on the
evolution of close binaries consisting of a low-mass component with a convective envelope filling its Roche lobe and a compact accreting component that is
a degenerate dwarf or a neutron star have been conducted by \citet{ver81},
\citet{rap83}, \citet{taa83}, \citet{pat84}, \citet{spr83}, \citet{tut84} and \citet{sar89}, among others.

The most likely angular momentum loss mechanism for W UMa systems
is magnetic braking. The observational evidence and theoretical
reasons have been presented for believing that low-mass W UMa
binaries can lose their
orbital angular momentum efficiently as a consequence of a
MSW from one or both components. The {\it
Einstein} soft X-ray observations and {\it IUE} ultraviolet
observations \citep{vai80,vil83,eat83,ruc83} showed that W UMa systems
are strong sources. This suggests surface activity of the kind
observed on our Sun, and so the presence of magnetic field. A
magnetic stellar wind would cause braking. In the absence of an
explicit theory for the angular momentum loss by MSW, we
employ a \citet{sku72} law, $V_{\rm rot}\approx 10^{14}\lambda
t^{-0.5}$, where $V_{\rm rot}$ is the rotating velocity in unit of
cm ${\rm s^{-1}}$, $t$ the age in seconds, and $\lambda$ a free parameter.
Observations give $\lambda=0.73$, according to \citet{sku72}, and
$\lambda=1.78$, according to \citet{smi79}. This estimates the rate
of angular momentum loss by a low-mass rotating star in a binary
system. Assuming that tidal coupling is perfect, so that the
wind-emitting star rotates synchronously with the orbital period,
and neglecting the spin angular momentum of each component
relative to the orbital angular momentum, we can take into account
the \citet{tut84} and \citet{ibe84} formalism for the loss of
angular momentum by a MSW in the following form
\begin{eqnarray}
\frac{{\rm dln}J}{{\rm d}t}&=&-9.6\ 10^{-15}\frac{R_{2}^{4}(M_1+M_2)^2}{\lambda^{2}A^{5}M_{1}}\ {\rm s^{-1}}\cr
       &=&-3.03\ 10^{-7}\frac{R_{2}^{4}(M_1+M_2)^2}{\lambda^{2}A^{5}M_{1}}\ {\rm yr^{-1}}
\end{eqnarray}
where $J$ is the orbital angular momentum of the system in ${\rm erg\cdot s}$,
$M_{\rm 1,2}$ are the masses of both components in solar masses $M_{\rm \odot}$,
$R_2$ is the radius of the donor in solar radii $R_{\rm \odot}$, and $A$ is the semimajor axis of the system in solar
radii $R_{\rm \odot}$. So far we have only indirect evidence of the magnetic
activity in W UMa-stars. Rapid period changes and disturbances in light
curves can be interpreted as being caused by big flares and spots \citep{bin77,
mul75,vant78}.

\section{Evolution into contact}

\begin{figure*}
\centerline{\psfig{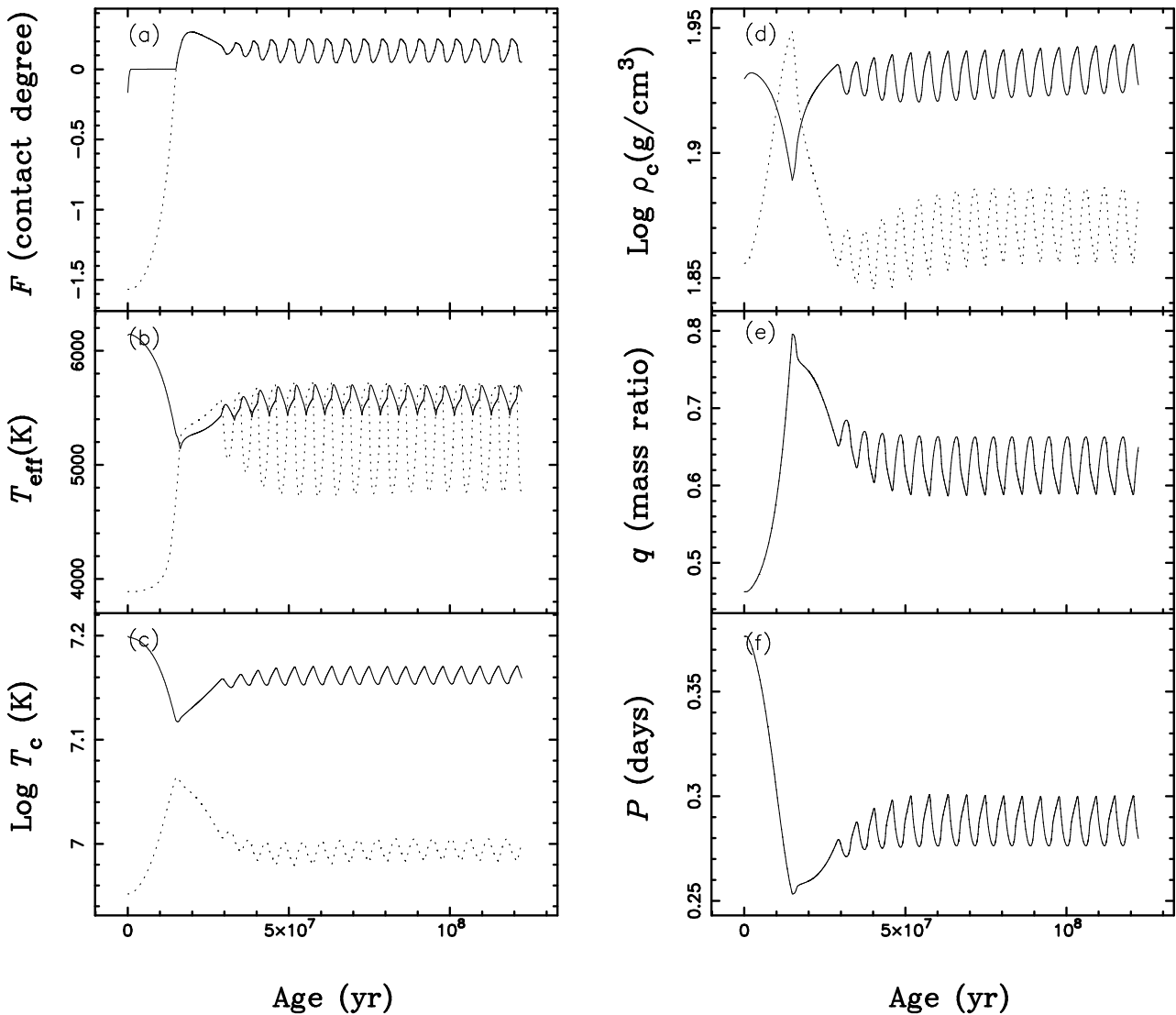}}
\caption{Without angular momentum loss and energy transfer due to
mass transfer, the evolution of some quantities [contact degree
$F$, temperature, central temperature, central density of the primary
(solid lines) and the secondary (dotted lines) , together with the mass
ratio and orbital period of the binary] as a function of age.}
\label{fig1}
\end{figure*}
\begin{figure*}
\centerline{\psfig{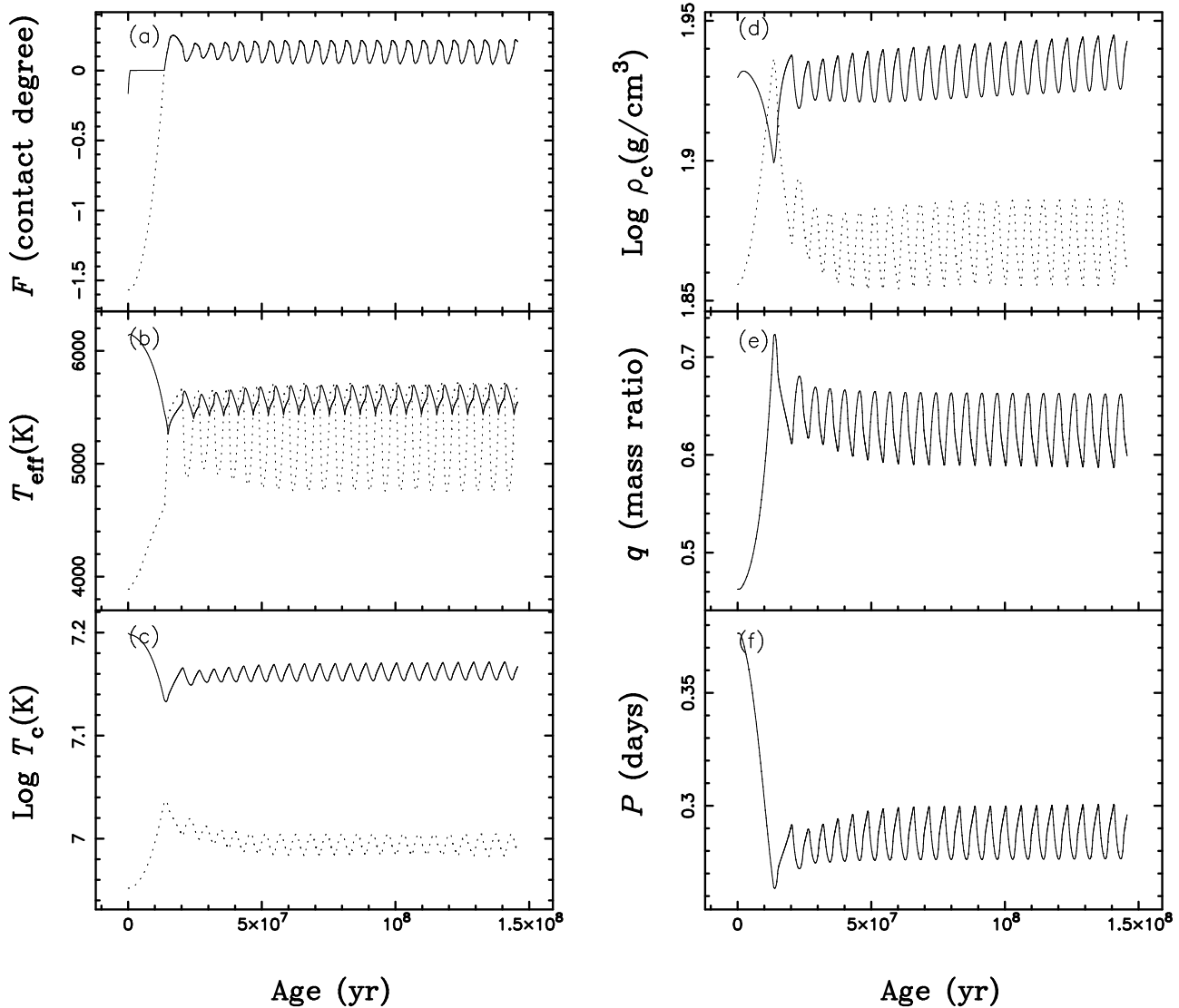}} \caption{
With the energy transfer due to mass transfer, but without AML, the same quantities as Fig. 1 against Age.}
\label{fig2}
\end{figure*}

\subsection{The initial model}

Our initial model consists of two zero-age main-sequence (ZAMS)
stars of Population I ($X=0.70, Z=0.02$) with a total mass of
1.8$M_{\rm \odot}$ ($1.23 + 0.57M_{\rm \odot}$) and an initial
mass ratio of 0.4634. The convective mixing length is twice the
pressure scale hight. We take an initial period of 0.3766 d and
orbital angular momentum, $J$, of $5.161\times10^{51}$ erg$\cdot$s
[${\rm log}(J/{\rm erg\cdot s}) =51.713$]. The initial separation
between two components is about $1.8574\times 10^{11}$ cm (i.e.
$2.67R_{\rm \odot}$). The initial model is a detached binary, and
the surface of the primary lies only a short way inside its Roche
lobe which it fills after about $7.3\times 10^5$ yr of nuclear
evolution.

\subsection{Contact evolution without angular momentum loss}

\subsubsection{Evolution without luminosity transfer due to mass transfer in semi-detached phase}

Our models in Paper I exhibit cyclic behaviour on a thermal timescale, with
a period of about $10^{7}$ yr. Part of the time of each cycle is spent in
a contact
phase and part in a semi-detached phase. At first we only consider the
luminosity transfer by circulation in the common envelope under the assumption
that no energy transfer due to mass transfer takes place in the semi-detached
phase, the system evolves in thermal cycles without loss of contact. The
changes of characteristic quantities (including the contact depth, surface
temperature, central temperature, and central density of the primary and
the secondary, together with the mass ratio and orbital period of the system)
during the cycles are shown in Figure 1.

As seen from Figure 1, beginning at the initial model, the primary fills
its Roche lobe after about $7.3\times10^{5}$ yr of nuclear evolution. It
evolves into a semi-detached state and it loses mass to its
companion. Since the addition mass to the secondary causes it to
expand, and its effective temperature and luminosity to increase, after
a total of about $1.5\times 10^{7}$ yr, the secondary has swollen to fill
its Roche lobe, so that the system evolves into a contact system. The contact
binary formed at this point has masses of 1.0 and 0.8$M_{\rm \odot}$, and a
mass ratio of about 0.8. Thereafter, the model
exhibits cyclic evolution on a thermal timescale
(about $6\times 10^{6}$ yr) without loss of contact. Part of time of each
cycle is spent in good thermal contact, with EW light curves and part in poor
thermal contact, with EB light curves. As seen
from Figure 1b, the ratio of the timescales characterizing the phases in which
the oscillating systems exhibit W UMa-type (EW) and $\beta$ Lyrae-type
(EB) light curves is $\tau_{\rm EW}/\tau_{\rm EB}<1$.

It is seen in Figures 1c and d that the central temperatures and central
densities of both components of the binary exhibit cyclic evolution. This
causes the cyclic evolution of the core luminosities of both
components of the binary.

\subsubsection{Evolution with luminosity transfer due to mass transfer in semi-detached phase}
\begin{figure*}
\centerline{\psfig{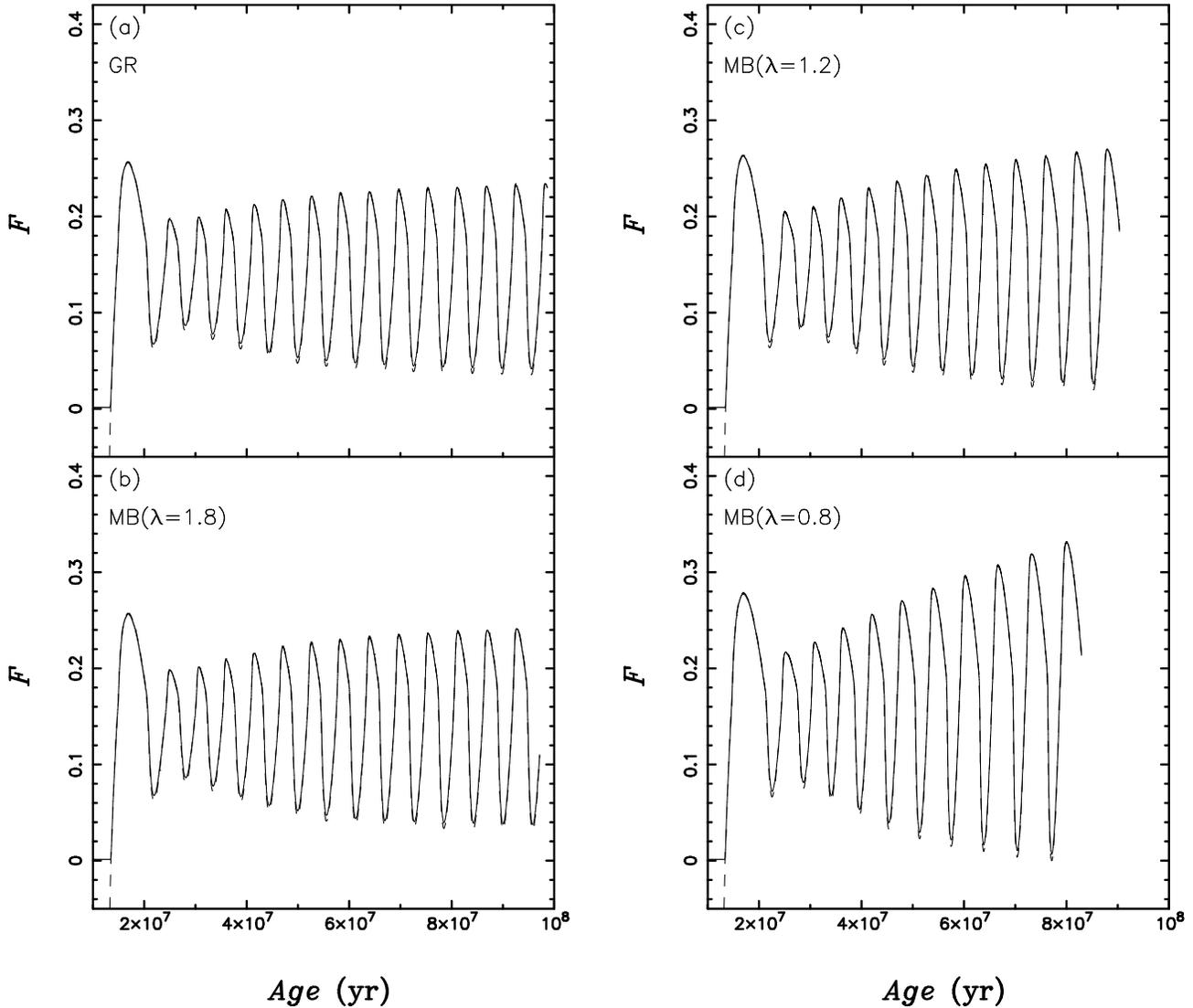}}
\caption{ Evolution of the contact degree of the binary with
angular momentum loss by GR or by MSW of the secondary (with the different values of
$\lambda$).} \label{fig3}
\end{figure*}
\begin{figure*}
\centerline{\psfig{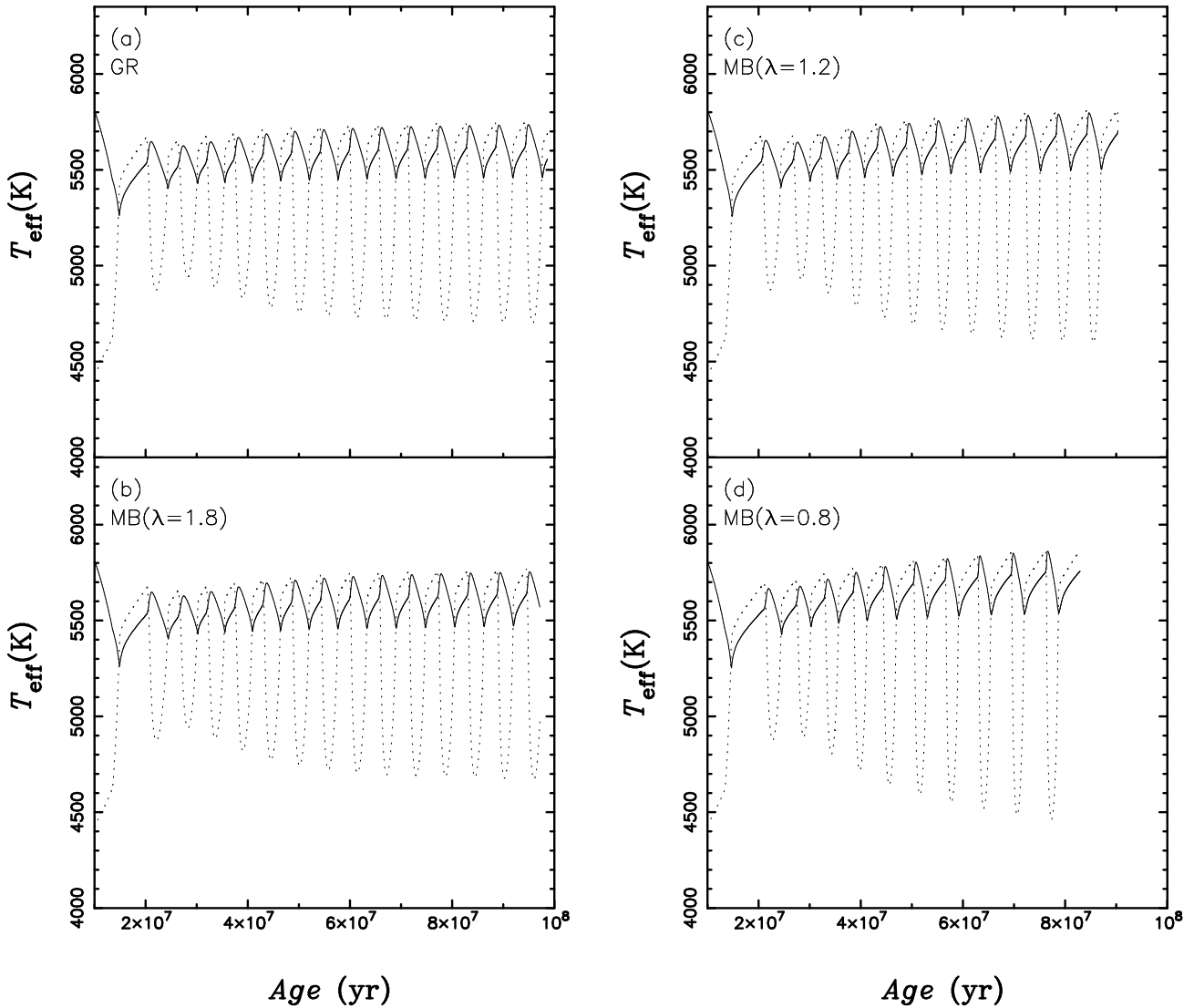}} \caption{
Evolution of the surface temperature ($T_{\rm eff}$) of both components of the binary with angular momentum loss
by GR or by MSW of the secondary (with the different values of $\lambda$).}
\label{fig4}
\end{figure*}
\begin{figure*}
\centerline{\psfig{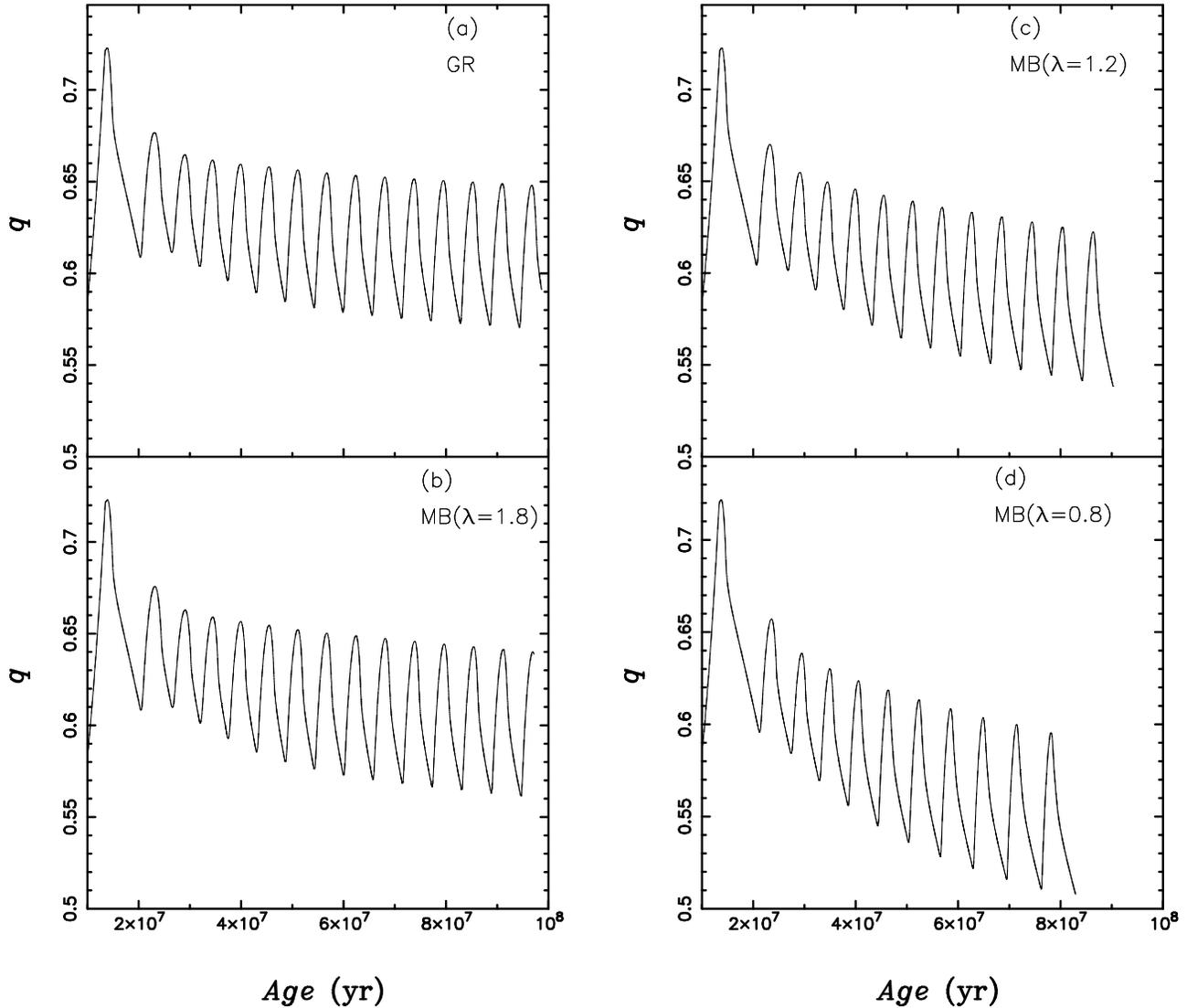}}
\caption{ Evolution of mass ratio ($q=\frac{M_2}{M_1}$) of the
binary with angular momentum loss by GR or by MSW of the secondary (with the different
values of $\lambda$).} \label{fig5}
\end{figure*}
\begin{figure*}
\centerline{\psfig{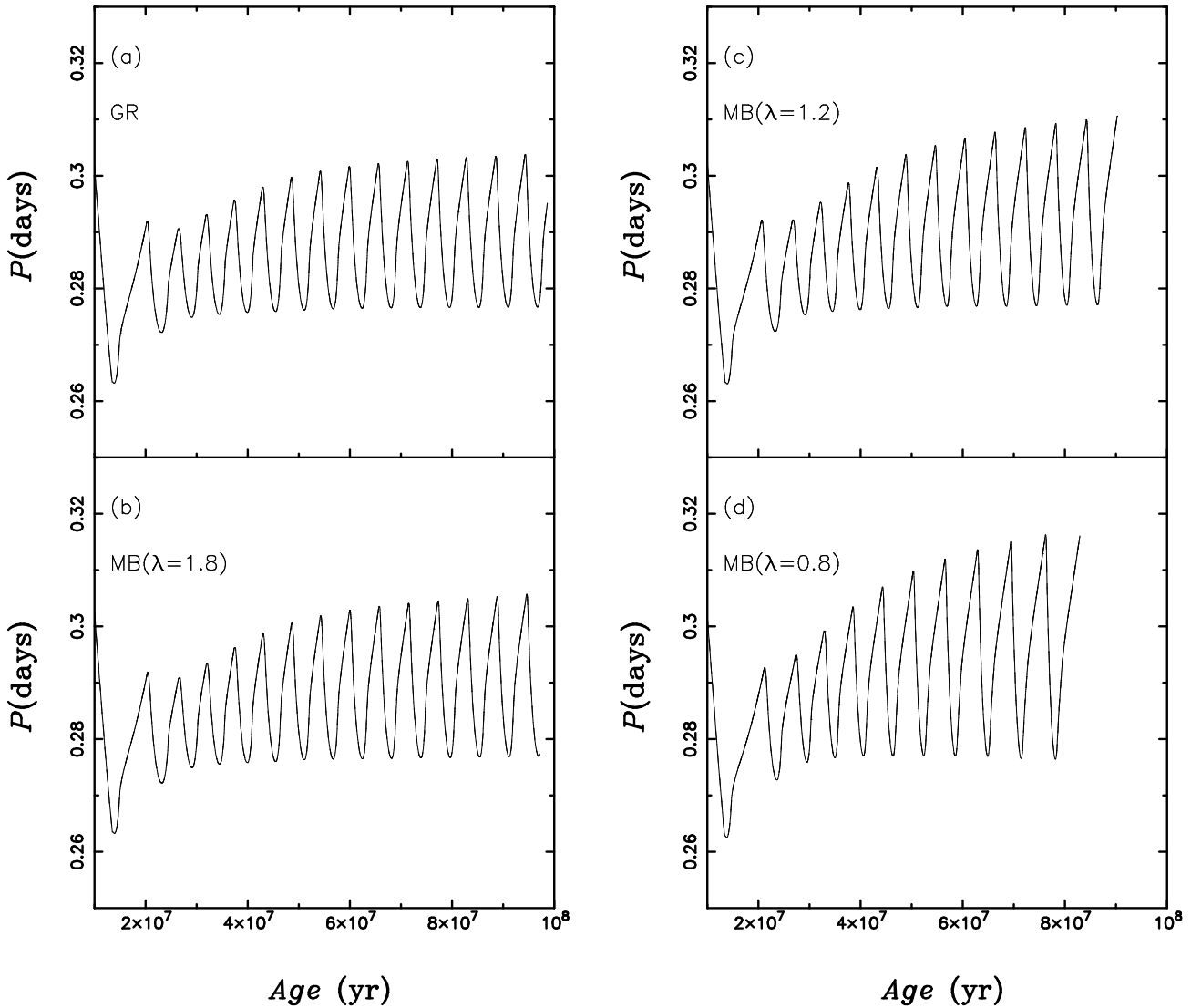}}
\caption{ Evolution of the orbital period ($P$) of the binary with
angular momentum loss by GR or by MSW of the secondary (with the different values of
$\lambda$).} \label{fig6}
\end{figure*}

We now consider the evolution of the binary system from the initial model
with luminosity transfer due to mass transfer included in the manner described
in Sect. 2.2.1. The system also evolves in thermal cycles
without loss of contact. The evolution of some characteristic quantities is
shown in Figure 2. It is seen in Figure 2 that the system evolves into a
contact binary from a detached one after a total of about $1.35\times 10^{7}$
yr of evolution from the main sequence and the contact binary formed at
this time has masses of about 1.045 and 0.755$M_{\rm \odot}$ with a mass
ratio of about 0.72. So this model spent shorter time
in the semi-detached phase than the model in which the luminosity
transfer caused by mass transfer had been neglected in the semi-detached state.
This is because the energy source at the secondary's atmosphere, provided by accreting
mass from the primary in the semi-detached phase, can hasten the expansion of the
secondary and so shorten the time spent in the semi-detached evolution.

It is seen in Figure 2b that the ratio the timescales ($\tau_{\rm EW}/\tau_{\rm EB}$) characterizing the phases
in which the oscillating systems exhibits W UMa-type (EW) and $\beta$
Lyrae-type (EB) is still less than 1.0. It is much smaller than
its observational value (5 to 6) derived by \citet{rah981} and even smaller
than a smallest value (13/5) derived by \citet{ruc02}. This suggests that the real W UMa systems may
suffer angular momentum loss. Therefore, we restrict the following discussions
to the possible evolutionary consequences of the models for W UMa systems with
angular momentum loss.

\subsection{Contact evolution with angular momentum loss}

\begin{figure*}
\centerline{\psfig{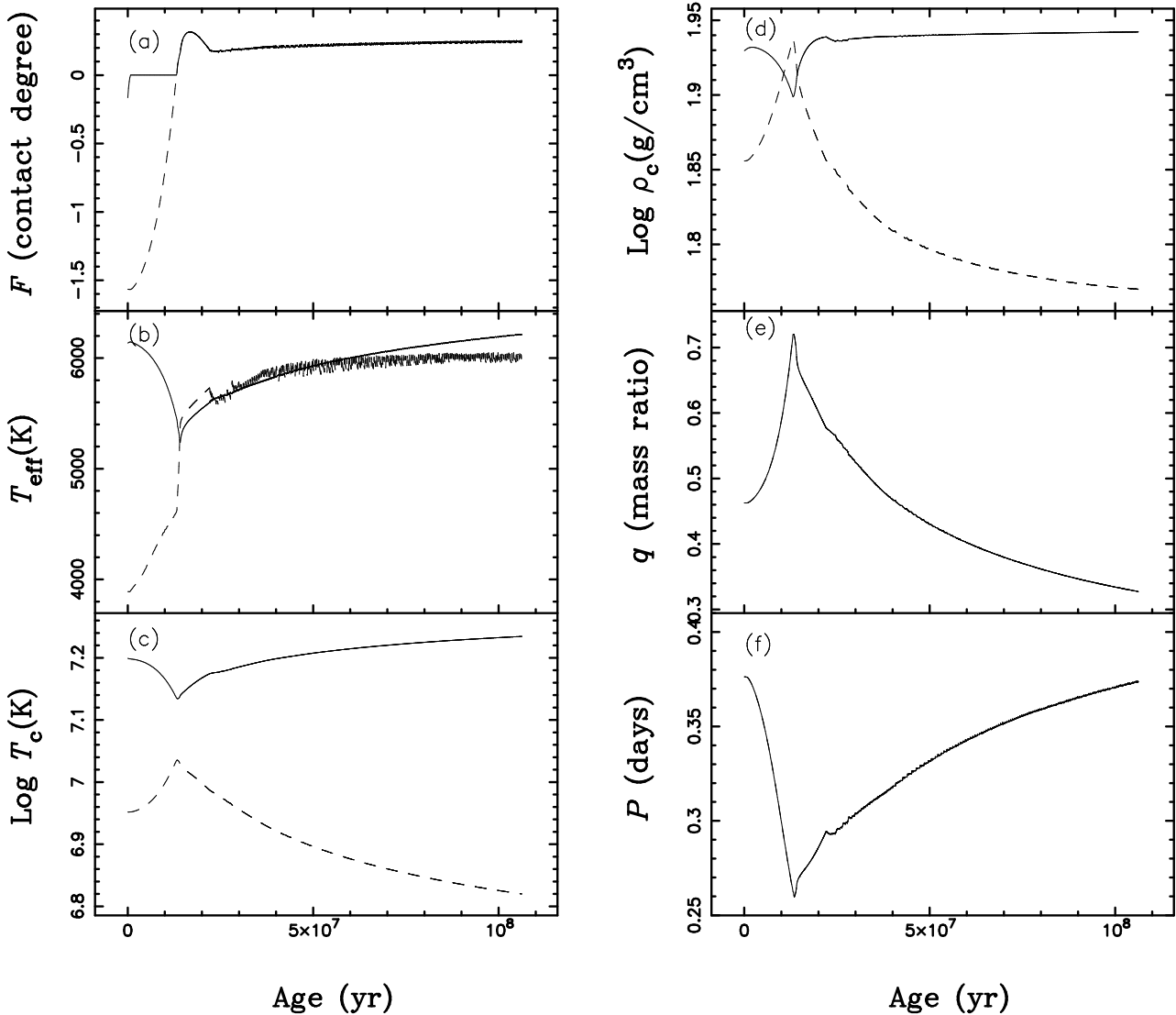}} \caption{
With angular momentum loss due to MSW ($\lambda=0.7$) of both components, the same quantities as figure 1 against Age. }
\label{fig7}
\end{figure*}
\begin{figure*}
\centerline{\psfig{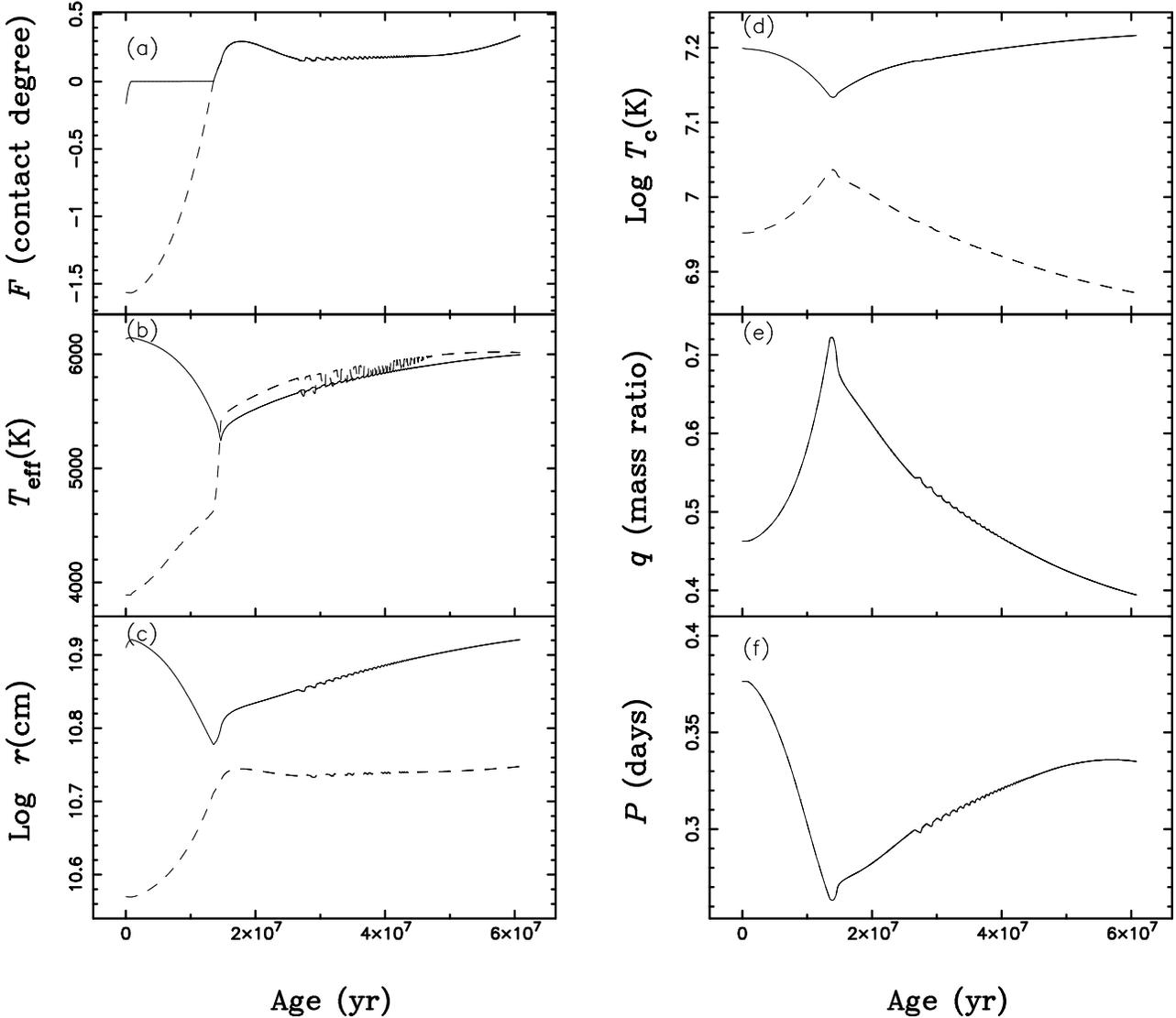}} \caption{
With angular momentum loss due to MSW ($\lambda=3.8$) of both components, the same quantities against Age. }
\label{fig8}
\end{figure*}
\begin{figure*}
\centerline{\psfig{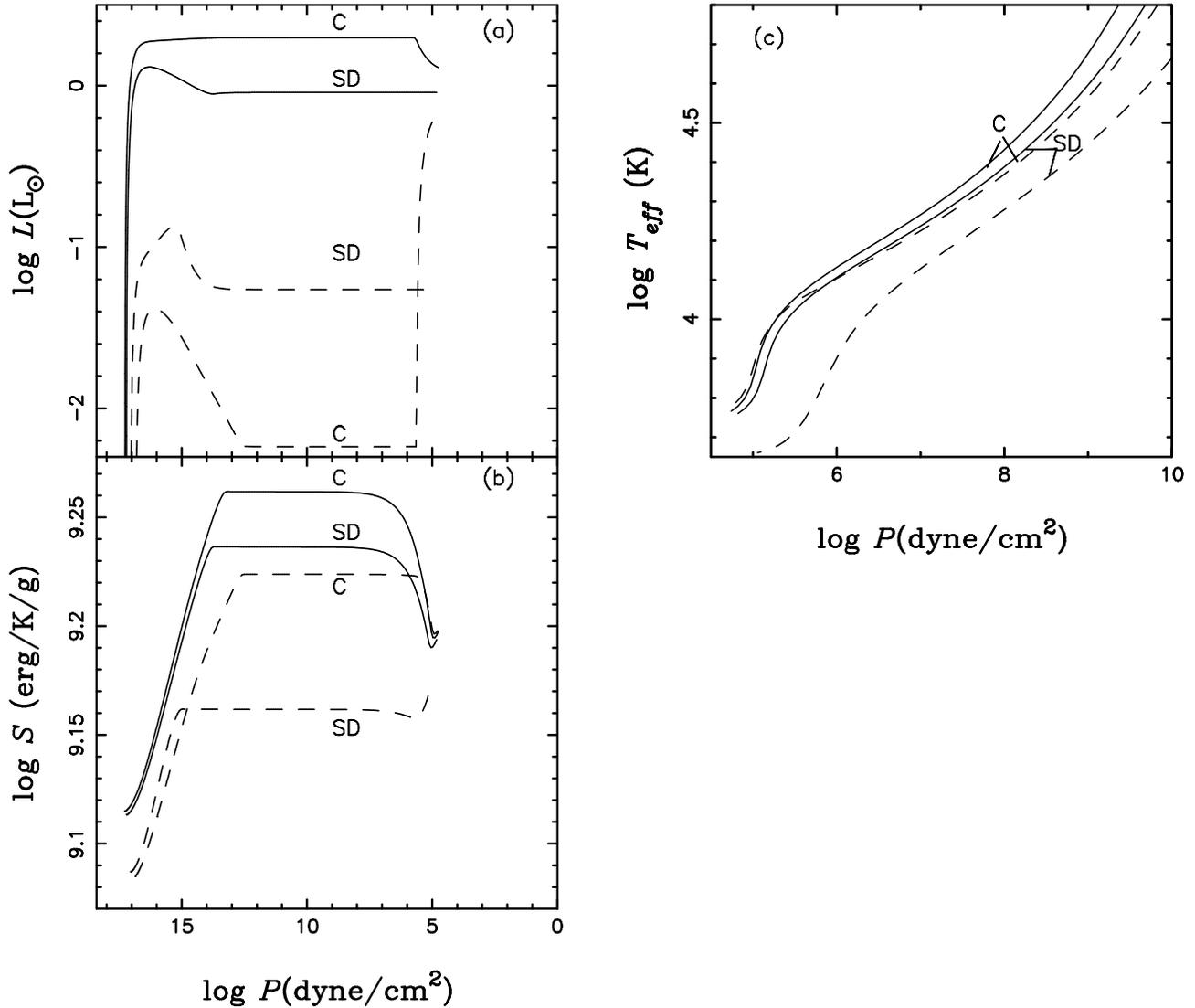}} \caption{
The thermal structure of the primary (solid lines) and the secondary (dashed
lines) during a phase of semi-detached evolution (SD) and of contact evolution
(C) with efficient energy transfer.}
\label{fig9}
\end{figure*}
\begin{figure*}
\centerline{\psfig{figure=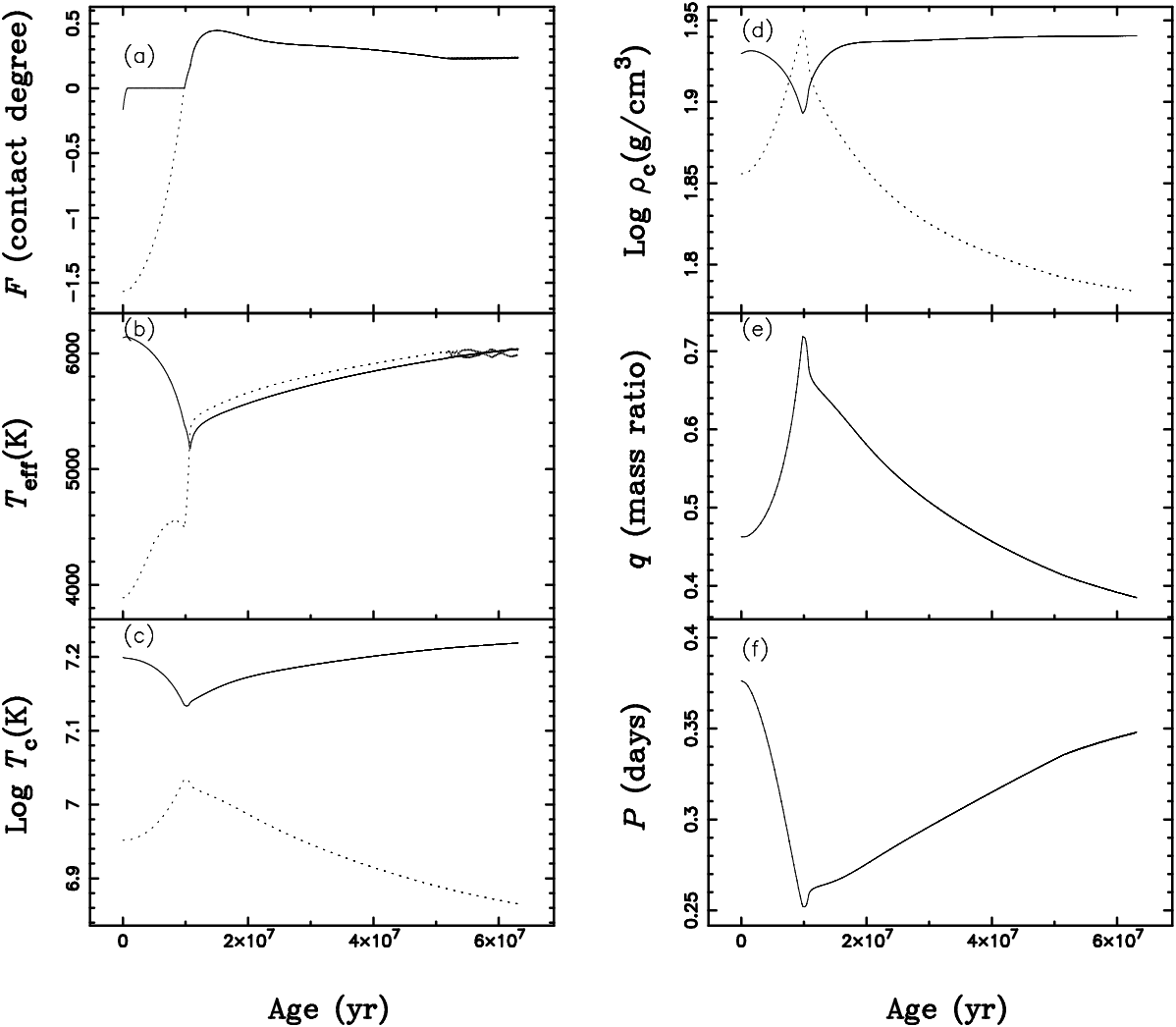,width=17cm,bbllx=580pt,bblly=109pt,bburx=79pt,bbury=686pt,angle=270}} \caption{
With a steady angular momentum loss, $\frac{{\rm dln}J}{{\rm d}t}\approx 1.6\ 10^{-9}{\rm yr}^{-1}$, the same quantities as figure 1 change with Age.}
\label{fig10}
\end{figure*}

\subsubsection{Evolution with angular momentum loss by GR}

Assuming that the angular momentum of W UMa-type systems is lost only by
gravitational wave radiation, we obtain a model for W UMa-type systems that
exhibits cyclic evolution without loss of contact as the models mentioned
above. The contact degree and surface temperature of both components,
together with the mass ratio and the orbital period of the system, are shown
in Figures 3a--6a. It is seen in Figures 3a--6a that the model exhibits cyclic evolution on
a thermal timescale (with a period of about $6\times10^6$ yr) without loss
of contact.

As seen from Figure 4a, the ratio of the timescales ($\tau_{\rm
EW}/\tau_{\rm EB}$) is still smaller than 1.0, this is still much smaller
than the value given by \citet{vil81} according to observations,
and even smaller than a smallest value (13/5) found by
\citet{ruc02}. This suggests that, although the gravitational
radiation has a significant influence on the evolution of W
UMa-type systems and other close binaries, it still can not
prevent the cyclic behaviour of the model for W UMa systems.
\citet{vil81} had concluded that if the timescale of angular
momentum loss for W UMa stars is shorter than the secondary's
thermal timescale (except at extreme mass ratios where the
primary nuclear evolution dominates) the system will soon fill
the outer critical surface and then probably coalesces into a single star,
if the timescale is longer than the primary nuclear timescale, the
effect will be negligible.
If the angular momentum loss rate falls
between these limits, it makes an important contribution to
the contact evolution. The rate of angular momentum loss
associated with gravitational radiation is
$10^{-11}$ to $10^{-10}\ {\rm yr}^{-1}$, corresponding to a timescale
of about $10^{10}$ to $10^{11}$ yr. It is clear that the gravitational radiation life-time is long compared with the nuclear timescale of the primary. Systemic angular
momentum loss due gravitational radiation must accelerate the evolution toward
more extreme mass ratios (see Figure 5a)
rather than lead to orbital collapse, so that
the effect of gravitational radiation on the cyclic evolution is almost
negligible. The most likely angular momentum loss
mechanism for W UMa systems is a magnetic stellar wind. There is considerable
observational and theoretical evidence that stars which possess convective
envelopes and, at the same time, rotation can amplify primordial magnetic fields,
emit a stellar wind and lose angular momentum as a consequence of the
interaction between the fields and the wind.

\subsubsection{Evolution with angular momentum loss by MSW}

Equation (12) gives a rate of angular momentum loss (via MSW) of a
close binary system in which the mass is transferred from the star 2 (with a
mass of $M_2$ and a radius of $R_2$) to star 1. This suggests
that the rate of angular momentum loss by MSW is directly related to
the direction of the mass transfer. Assuming that the mass is
always transferred from the secondary to the primary in W UMa
systems, and taking $\lambda$=1.8, 1.2, or 0.8, which are in
its observational range, we investigate the influence of the
parameter $\lambda$ on the evolution of the models for W UMa systems.
In these cases the evolution of contact degree and the surface temperature
of both components, together with the mass ratio and the orbital period of
the system, are also plotted in Figure 3b--6b, c, and d, respectively.

It is seen in Figures 3--6 that our models still exhibit cyclic
behaviour without loss of contact on the thermal timescale of the
secondary. This is similar to the case of constant angular
momentum with the exception that the cycles are slowly shifted
towards more extreme mass ratios. And that the smaller the
parameter $\lambda$ is taken to be (i.e. the higher the rate of
angular momentum loss is assumed to be), the longer is the period
of cyclic evolution, and the larger is the ratio of the timescales
characterizing EW and EB phases ($\tau_{\rm EW}$ and $\tau_{\rm
EB}$). If we take $\lambda=0.8$, the ratio of the timescales
($\tau_{\rm EW}$ and $\tau_{\rm EB}$) has risen up to 2 to 3
which is consistent with the result obtained by \citet{ruc02}.
Therefore, the most likely angular momentum loss mechanism for W
UMa systems is MSW, and the value of $\lambda$ is of about 0.8 for
W UMa systems if the mass is transferred from the secondary to the
primary in W UMa systems. Meanwhile, as seen from Figures 3--6, the
mean mass ratio in a cycle becomes smaller and smaller as the
evolution of the system proceeds, in any case with angular momentum
loss, the mean orbital period of the system becomes longer and
longer. I.e. the system drifts to smaller and smaller mass ratios
and to longer and longer periods along contact branches of the
cyclic evolution. So W-type W UMa systems evolve
into A-type W UMa systems with low total masses. Since the systems
with $P<0.35$ d are apparently not observed to have EB-type light
curves, the fraction of time spent in the broken-contact phase
or in poor thermal contact phase is uncertain and possibly small.
This suggests that the poor thermal contact or semi-detached state
probably appears very rarely (or never appears) during the
evolution of the real W UMa systems. In effect, the loss of
angular momentum due to MSW is used here to gently squeeze the
systems and thus prolong the contact phase of the TRO model to
such a degree that the broken-contact or poor thermal contact
phase appears very rarely (or never appears).

If the mass is transferred from the primary to the secondary, the rate of
angular momentum loss should be expressed as
\begin{equation}
\frac{{\rm dln}J}{{\rm d}t}=-3.03\ 10^{-7}\frac{R_{1}^4(M_1+M_2)^2}{\lambda^2 A^5M_2}\ {\rm yr}^{-1}.
\end{equation}
We take $\lambda=1.8$ which is the largest value obtained by the observations
of single stars, and corresponds to the smallest rate of angular momentum loss due
to MSW for single stars, and find that the model soon fills its outer critical
Roche lobe after a total of about $2.5\times 10^7$ yr of nuclear evolution. The
system loses its mass and angular momentum through the outer Lagrangian point.
Then it probably coalesces into a single star. Since the direction of mass transfer
may change with time, we take $\lambda=1.8, 1.2$, and $0.8$,
respectively, and calculate the rate of angular momentum loss via Eq.
(12) when mass is transferred from the secondary to the primary, otherwise via
the Eq. (13). In these cases, the models rapidly drift to extreme mass ratios,
and they cannot be in good thermal contact when the mass ratio of the models
becomes sufficiently small. Because, once the mass ratio of the system is very
small, the mass is transferred from the primary to the secondary when the model
is in poor thermal contact, then the rate of angular momentum due to MSW of
the primary is very large, and the model evolves towards good thermal contact
rapidly, but the direction of mass transfer has been reversed before the models
are in good thermal contact, and the rate of angular momentum loss due to MSW
of the secondary is too small to further force the model to evolve towards deep
contact. Therefore, once the mass ratio of the models
is sufficiently small the model is always in poor thermal contact even
if we take the parameter, $\lambda=0.7$ (see Figure 7) although
the models exhibit thermal cyclic behavior with a short period and
a small amplitude.

In effect, W UMa systems are surrounded by a common envelope and the loss
of angular momentum is probably not related to the direction of the mass
transfer and may be simultaneously caused by MSW of both components.
We take $\lambda=3.8$, and the loss rate of angular momentum of the model
is taken as the sum of loss rates predicted by Eqs (12) and (13). The evolution
of some quantities is shown in Figure 8. The thermal structure of the
two stars of the binary ($\lambda=3.8$) during a phase
of contact evolution and of semi-detached evolution is shown in
Figure 9. It is seen in Figure 8 that our model almost maintains
an essentially constant depth of contact, rather than cycle in evolution,
and that the system steadily drifts to smaller and smaller mass ratios.
This suggests that the model evolves steadily towards a system with the extreme
mass ratio (see Figure 8e). Meanwhile, if we assume a somewhat smaller value
of $\lambda$ (i.e. a larger rate of angular momentum loss),
the system soon reaches the outer critical Roche lobe. It then loses
its mass and angular momentum through the outer lagrangian point, so that it
probably coalesces into a rapidly rotating single star. On the other hand, if
the parameter $\lambda$ is assumed to be somewhat larger, the system exhibits
cyclic evolution, and the subsequent evolution of the system is similar to the
case of constant angular momentum with the exception that the cycles are slowly
shifted towards more extreme mass ratios. In this case, the parameter $\lambda$
is much larger than the value of similar single or non-contact stars.
This suggests that the magnetic braking in W UMa systems is weaker than
that in the similar single stars or components of non-contact binaries.
The observations have shown that both in X-rays \citep{cru84,vil84} and
in the radio \citep{hug84,hug85} the W UMa systems are definitely less
active than are similar single stars or components of non-contact binaries.
The lower coronal activity in W UMa systems is probably caused by the internal
rotation of their components or by their convective zone which is thinner than
indicated by their surface temperatures \citep{ruc86}. The relative weakness of
coronae in W UMa systems directly tells us that the largest-scale
magnetic structures are smaller in W UMa systems and contain less plasma in
them than in other similar but non-contact stars. Therefore, the
magnetic-braking effects in W UMa systems are weaker than in other similar
but non-contact stars. Meanwhile, we can see from Figure 8b that the surface
temperature of the secondary is unstable, so that its luminosity is unstable.
This is consistent with the properties of W-type W UMa systems with
variable light curves. Moreover, we can see from Figure 8c that the radius
of the primary is expanding rapidly while the radius of the secondary appears
to be essentially constant although the model is drifting to smaller
and smaller mass ratios. This is very similar to the result predicted
by observations \citep{wan95}. This suggests that W phenomenon is likely caused by
the expansion of the primary (Paper I) because the expansion of the radius
of the primary depletes a part of the core luminosity so that the
temperature of the secondary exceeds that of the primary.

As seen from Figure 9, the outermost layers of the common envelopes of both
components are so similar that the difference is slight. Meanwhile, we can
see in Figure 9b that the profile of the specific entropy in the surface
layers of the secondary is very different from the result
given in Paper I. It clearly rises up to a higher value in the surface layers
of the secondary because of the energy transfer due to mass transfer.
Figure 12 shows the distribution of the specific entropy in both components as
a function of the mass (Figure 12a) and the radius (Figure 12b) during a phase
of semi-detached evolution and of contact evolution with efficient energy
transfer. The semi-detached model has the same total mass and mass
ratio as the contact one. The flat parts in outer envelopes of the
components indicate the convective regions. It is seen in Figure 12
that the convective envelope of the primary in the contact system
is slightly thicker than that of the primary in the semi-detached one and
the convective envelope of secondary in contact system is much thinner
than that of the secondary in the semi-detached one (see Figure 12b),
but the total mass contained in the convective envelopes of the contact system,
which is directly related to the dynamo action \citep{hur02}, is much less
than that
contained in the envelopes of the semi-detached one (see Figure 12a). We
conclude that this is caused by luminosity transfer between the two components.
Because part of the luminosity in the envelope of the primary
is transported to the envelope of the secondary in the contact state, the
temperature of the envelope of the primary decreases and the temperature of
the envelope of the secondary increases. The decrease in the temperature
of the envelope of the primary would then drive the radiative temperature
gradient, $\nabla_{\rm r}$, to increase and thence lead part of the
radiative zone to be convective so that the the convective
envelope of the primary becomes thicker in the contact state.
The increase in the temperature of the envelope of the secondary would drive
the radiative temperature gradient, $\nabla_{\rm r}$, to decrease and
thence lead part of the convection zone to be in radiative equilibrium.
As a result, in contact systems, the mass in the convective envelope of the
primary increases
by about 0.01 $M_{\rm \odot}$, and the mass in the convective envelope of the
secondary decreases by about 0.08 $M_{\rm \odot}$ in comparison with
in the semi-detached one (see Figure 12a). Since the
decrease of the mass in the convective envelope of the secondary
is much more than the increase of the mass in the convective envelope of the
primary, the total convective mass in the envelope of the contact
system is less than that in the envelopes of the semi-detached one. As a result
of the less total mass contained in the convective envelopes of the W UMa
systems, because of energy transfer, W UMa systems show lower activity than
non-contact stars.

We also investigate the evolution of the models which lose
angular momentum steadily at the different rates, and find that the system
can keep in shallow contact (with a degree of about 20\%, see Figure 10) by
a hypothetical angular momentum loss $\frac{{\rm dln}J}{{\rm d}t}\approx 1.6\ 10^{-9}{\rm yr}^{-1}$, on the average, corresponding to a timescale of about $6\ 10^8$ yr which
is in agreement with the result obtained by \citet{rah981}. If we assume a
somewhat larger rate, the system soon reaches the outer critical surface.
This was confirmed with a steady loss of angular momentum with a rate
that was slightly larger than the critical loss rate (see also Robertson \&
Eggleton, 1977). After reaching the outer critical surface, the system
loses its mass and its angular momentum through the outer Lagrangian points,
so that the system probably coalesces into a single star \citep{web76,rah981}.
On other hand, if the assumed loss rate is smaller than the
critical rate, the system exhibits cyclic evolution with alternating good
contact and poor contact phases on the secondary's thermal timescale.
In this case the subsequent evolution is
similar to the case of constant angular momentum with the exception
that the cycles are slowly shifted towards more extreme mass ratios.

\subsection{The evolution in the period-colour diagram}

\begin{figure}
\centerline{\psfig{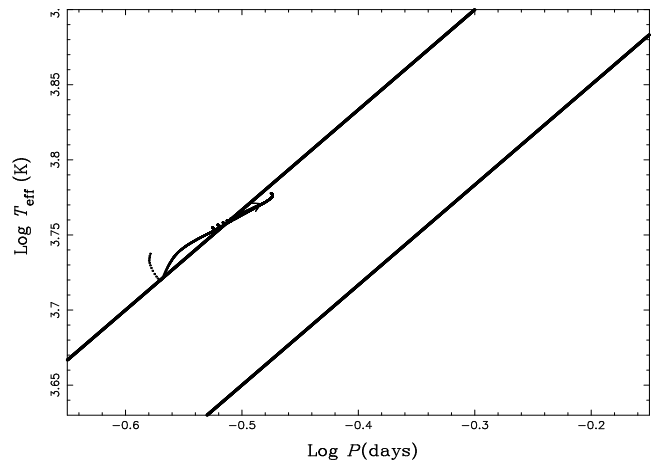}}
\caption{
The evolution of the massive component of the binary in the period-colour diagram.}
\label{fig11}
\end{figure}
\begin{figure*}
\centerline{\psfig{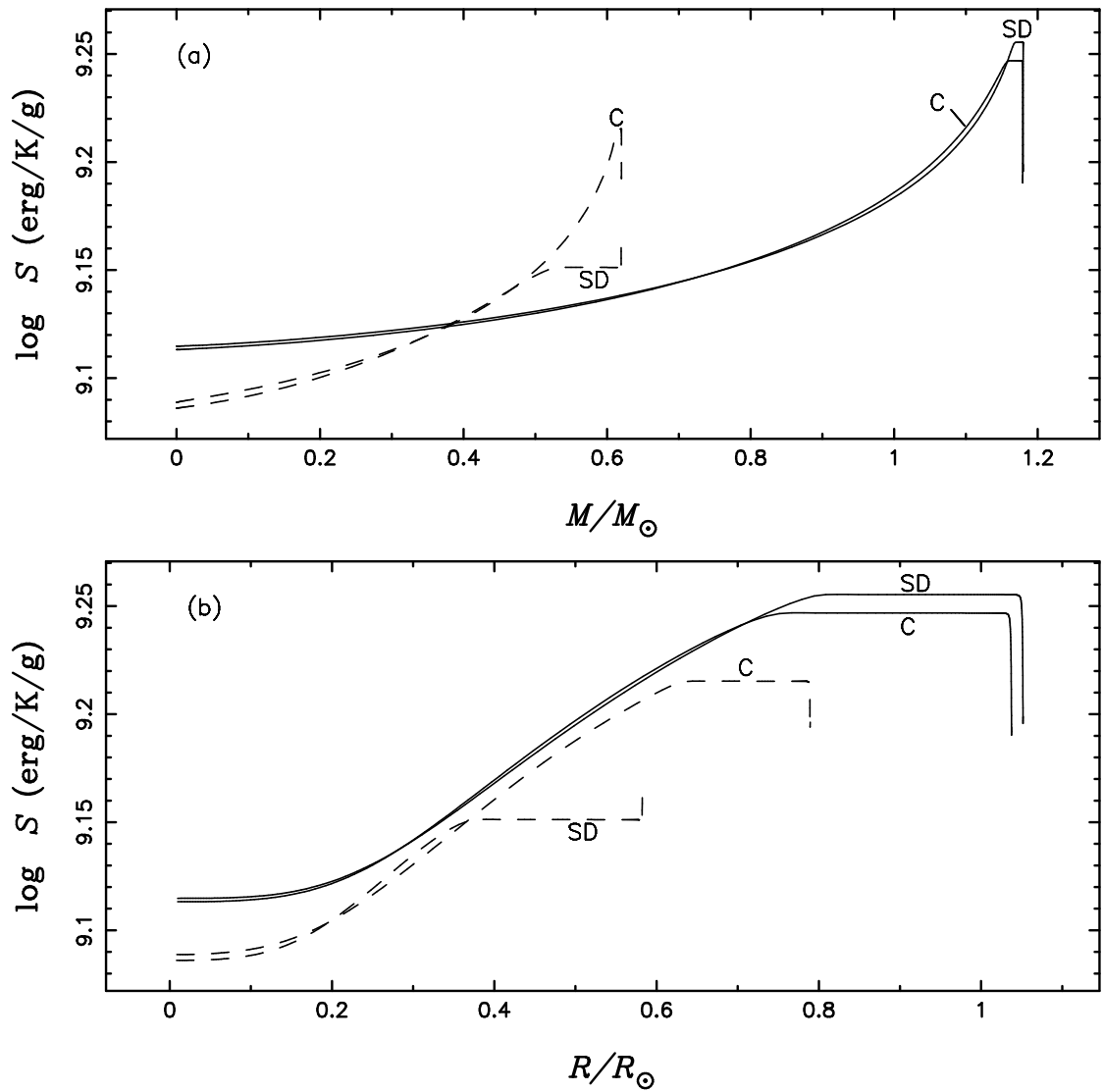}} \caption{
The distribution of the specific entropy of the primary (solid lines) and the secondary (dashed
lines) against mass (a) and radius (b) during a phase of semi-detached
evolution (SD) and of contact evolution (C) with efficient energy transfer.}
\label{fig12}
\end{figure*}

As shown by \citet{egg61,egg67}, the observed W UMa systems are located
in a strip of the period-colour diagram, the two observed boundaries of the period-colour diagram are written by \citet{kah02b} as
\begin{equation}
1.5{\rm log} T_{\rm eff} - {\rm log} P=5.975...6.15,
\end{equation}
where $P$ is the period in days. This region is limited by the solid lines in
Figure 11. Also shown is the massive component of
our model [${\rm log}(J/{\rm erg\cdot s)}=51.713$, and $\lambda=3.8$].
As seen from
Figure 11, our model is in a good agreement with the period-colour
relation for observed W UMa systems. Moreover, our model is at least
similar to Rahunen's (1981) one.

\section{Discussion and Summary}

In semi-detached phase, the mass is transferred from the primary to the
secondary, and the energy (including the gravitational energy, thermal energy,
and radiative energy) in the transferred mass is also transferred
to the surface of the secondary. The effect of the
accreting matter on the radii of the accretors is related to their envelopes
being radiative or convective, and calculations involving mass accretion on to
a star with thick convective or fully convective envelope \citep{why85}, have
shown that the accreting star shrinks (under the assumption
that the entropy of the accreting matter is the same as the surface entropy
of the original star). However,  \citet{sar89} thought that the newly
accreting matter has substantially higher entropy than the original star and so
the photosphere of the secondary in close semi-detached binaries is heated by
the infalling matter. Due to this process, the accreting secondary should
significantly increase in radius \citep{pri85}. That is to say, the energy
source at secondary's atmosphere provided by the accreting matter
from the primary hastens the expansion of the secondary and so shortens the
time spent in the semi-detached evolution. Therefore, the luminosity transfer
due to mass transfer plays an important role in the origin and evolution of
contact binaries. Meanwhile, if the energy transfer due to mass transfer is
applied in the surface
layer of the secondary as a surface condition of the secondary, we can see that
this energy source has a significant influence on the
thermal structure in the surface layers of the secondary (see figure 9b).
As seen from figure 9b, the entropy profile in the surface layers of the
secondary in the semi-detached phase rises. It is very different from
the result in Paper I (see Paper I's Figure 6), and the difference arises from
the luminosity transfer due to mass transfer.

W UMa systems have been divided by \citet{bin70} into A-type and W-type
systems according to whether the primary minimum in the light curve is a
transit or an occultation. This suggests that the temperature of the more
massive component is higher in A-type systems while the temperature of the
less massive star is higher in W-type ones, i.e, the so-called W phenomenon.
It is seen in Figure 8c that the radius of the primary is expanding
in good contact while the secondary almost maintains an essentially
constant radius in low-mass W UMa systems. Since the expansion of the
primary depletes a part of the core luminosity, the temperature
of the secondary exceeds that of the primary. This suggests that
W-phenomenon is caused by expansion of the primary (see Figure 8b and c).
It is similar to the statistical result derived by \citet{wan95} from the
observational data.

Our model exhibits cyclic behavior without loss of contact,
with a period of about $6\times10^{6}$ yr if we do not consider
angular momentum loss from the system. And the ratio of the
timescales characterizing the phases in which the oscillating
systems exhibit W UMa-type (EW) and $\beta$ Lyrae-type (EB)
light curves is $\tau_{\rm EW}/\tau_{\rm EB}<1.0$. This is much
smaller than its observational value (5 to 6, Rahunen 1981 and
2 to 3, Rucinski 2002). Meanwhile, all constant angular momentum
loss models need a formation mechanism which can produce unequal
components because, with constant angular momentum, it is not
possible to cover a large range of mass ratios
\citep{vil81,rah981}. This suggests that the real W UMa systems
should suffer angular momentum loss during their evolution. The
rate of angular momentum loss associated with gravitational
radiation is about $10^{-11}$ to $10^{-10}\ {\rm yr}^{-1}$ for W UMa
systems \citep{rob77, web75,rah981}. This is two orders of
magnitude smaller than the rate of the hypothetical angular
momentum loss needed to keep the systems in good thermal contact.
The effect of gravitational wave radiation on the cyclic evolution
is so small that the model with angular momentum loss due to GR is
not consistent with the observations of W UMa systems.
The most likely angular momentum
loss mechanism is MSW. In fact, there are some indications of
strong magnetic activity in W UMa systems. \citet{vant79} thinks
that magnetic braking may control the evolution of solar-type
contact binaries. Eqs. (12) and (13) give the rate of angular
momentum loss by MSW of the secondary and the primary,
respectively, and the observations give the value of parameter,
$\lambda=0.73$ to 1.78 for single stars. According to Eqs. (12)
and (13) the rate of angular momentum loss caused by MSW is
directly related to the direction of the mass transfer in a close
binary system. If we assume that the mass is always transferred
from the secondary to the primary, and take $\lambda=0.8, 1.2$,
and 1.8, respectively. We find that our models exhibit cyclic
evolution which is similar to the case of constant angular
momentum with the exception that the cycles are slowly shifted to
the more extreme mass ratio, and the larger the rate of angular
momentum loss is assumed to be, the more rapidly the model evolves
towards the extreme mass ratios, and the larger is the ratio of the
timescales ($\tau_{\rm EW}/\tau_{\rm EB}$). If the parameter,
$\lambda$, is taken as 0.8, the ratio of the timescales
($\tau_{\rm EW}/\tau_{\rm EB}$) is about 2 to 3 which is in
agreement with a observational value \citep{ruc02}. Meanwhile, we
can see from Figures 3--6 that the mean mass ratio of a cycle
becomes smaller and smaller and the mean period becomes longer and
longer as the evolution of the system proceeds. This indicates that
it slowly evolves towards a system with an extreme mass
ratio and a long period. If the mass is assumed to be
transferred from the primary to the secondary, even if the
parameter,$\lambda$, is taken to be 1.8, the model soon fills
its outer Roche lobe. The star would then lose its mass and angular
momentum through the outer Lagrangian point, and the system would
soon evolve into a single star. If the direction of the
mass transfer is changed with time, and the angular momentum loss
of the system is caused by MSW of the secondary when the mass is
transferred from the secondary to the primary or by MSW of the
primary when the mass is transferred from the primary to the
secondary, no matter how the parameter $\lambda$ is chosen
between 0.73 and 1.78, the model cannot be in good
contact when it evolves to a system with a sufficiently
small mass ratio.

Our models for low-mass W UMa systems show that the magnetic braking
effects in W UMa systems are likely to be weaker
than that in other similar but non-contact stars. This is in agreement with
observations. Although there were both observational and theoretical
indications \citep{rob74,kno82} of a power law ($B=B_0 \Omega^{\delta}$, where
$\Omega$ is the Keplerian angular velocity, $B$ the average poloidal field,
with the most probable value of $\delta\approx 1$). However, observation
has shown that the strengths of the chromospheric and transition-region
emissions increase very slowly with the increase of the inverse to
Rossey number $\tau/P$ \citep{ruc86,ste95}. This can be
explained as a result of the total saturation of stellar surfaces by small
active regions. Meanwhile, the observations have shown that, both in X-rays and
in the radio, the W UMa systems are definitely less active than are similar
single stars or components of non-contact binaries \citep{ruc86}. The lower
coronal activity in W UMa systems is likely caused by the internal rotation
of their components or because their convective zone is thinner than
indicated by their surface temperatures. The observations
directly tell us that in W UMa systems the largest-scale magnetic structures
are smaller and contain less plasma in comparison with the
similar but non-contact stars so that the efficiency of the angular momentum
loss in W UMa systems may be lowered in comparison with non-contact stars of
the same $\tau/P$. Figure 12 shows the distribution of the specific entropy
of both components during a phase of a semi-detached state and of a contact
state. It is seen in Figure 12 that the convective envelope of the primary
in the contact
system is slightly thicker than that in the similar semi-detached one and the
convective envelope of the secondary in the contact state
is much thinner than that in the semi-detached state, but the total mass
contained in the convective envelopes of contact systems, which is related to
dynamo action \citep{hur02}, is less than that contained
in the convective envelopes of the semi-detached stars because of energy
transport from the primary to the secondary (see Figure 12). This is in a good
agreement with the prediction of observations mentioned above.
As result of the less total mass
in convective envelopes of W UMa systems, W UMa systems show the lower activity
in comparison with the single or non-contact stars.

Our models indicate that W UMa systems evolve into contact binaries with the
extreme mass ratios or even into single stars if they suffer angular momentum
loss due to MSW. \citet{vil81} thought that W UMa systems most probably evolve
into single stars, but the way (and the timescale) in which this evolution
proceeds is not clear. However, it is argued that due to a longer timescale
to reach a contact, and also to the longer timescale of contact evolution,
the ultimate appearance of low-mass binaries after merging of both components
will be distinctly different from that of massive binaries.
After an initial mass exchange
of a massive contact system, it is very likely that the main-sequence (MS)
lifetime of a more massive component becomes shorter than the duration of
the contact phase of evolution, and that the coalescence results in the
formation of a fast-rotating giant, similar to FK Com. In the case of
low-mass binary, the MS lifetime of a more massive component may still
be longer than the duration of the contact phase. That would result
in the formation of a single fast-rotating MS star, similar to
single blue straggler \citep{ste95}. The contact systems with
most extreme mass ratios (few known due to the low probability of eclipses)
are important for testing models of contact binary evolution.
But most W UMa stars with low mass ratios (such as RR Cen and $\epsilon$ CrA,
see Vilhu 1981) seem to have an evolved primary and both components should
deviate from thermal equilibrium.

Although we can keep our models in good contact under some assumptions,
there is still one observational point which seems to be in
contradiction with the scenarios presented above, because it produces a
peculiar mass ratio distribution showing a large excess of the systems with
small mass ratios. This is due to the fact that the mass transfer occurs on
the thermal timescale of the secondary, which grows strongly as the secondary
mass and therefore as the mass ratio decreases. Thus systems with large mass
ratios initially evolve in less than $10^{8}$ yr towards mass ratios less
than about 0.4 (see Figures 8 and 10). The scenarios in which
contact-binary evolution continues into a single star on the secondary
thermal timescale produce a rather steep period-distribution. It can be
shown quite easily that the resulting
mass ratio distribution does not agree with the mass ratio distribution
deduced from observation \citep{vant78,vil81}. Meanwhile, W UMa-type
contact binaries are found in stellar aggregates of a widely differing age,
in very old clusters like NGC188 and in moderately old ones like M67 and
Praesepe. This cannot be interpreted by these models.
The decrease in the coronal (X-rays and radio) activity of contact binaries,
which is indeed observed \citep{cru84,vil84,hug84}, is used as an argument
that the AML efficiency in contact is relatively low and the contact stage is
considerably prolonged relative to adjacent stages. This small modification to
the AML models is capable of explaining why so many different contact binaries
are observed in old systems like NGC188. Therefore, a feedback mechanism is
crucial to the success
of these models. \citet{vil81} envisaged a situation in which increasing depth
of contact causes increased mixing in the common envelope which tends to bury
the strong surface magnetic field. The magnetic braking is then weakened,
leading to a decreased loss rate of angular momentum and decreased depth
of contact again. It is clear that this picture is very crude and
speculative. On the observational side one should try to measure the
magnetic fields and to study the chromospheres and coronae more closely
in the UV- and X-ray region. On the theoretical side one should clarify the
differential rotation and dynamo action in rapidly rotating stars in connection
with the circulation pattern in the common envelope, a task which is not
easy \citep{vil81}. \citet{ruc82} explained the period gap at periods just
longer than those for contact systems by a phase of very rapid magnetic braking
in detached binaries which ends when the stars come into contact. However,
it is unclear that this mechanism is sensitive enough to maintain an
essentially constant depth of contact.

The spin angular momentum of both components and the loss of the
system's mass have not been considered in our present work. Indeed the spin
angular momentum of both components is not small compared with the orbital
angular momentum when the mass ratio becomes extreme \citep{vant79}.
If the spin angular momentum of both components is considered,
angular momentum loss required to spin up the stars to keep them
in corotation as the orbit shrinks, so the spin angular
momentum of both components increases and the orbital angular momentum
of the system decreases, the system might then suffer Darwin's
instability (i.e., the tendency for an orbit to desynchronize if
the spin angular momentum of both stars is more than a third of the orbital
angular momentum, Eggleton \& Kiseleva-Eggleton 2001) when so much angular
momentum has been removed that there is no longer sufficient in the orbit
to bring the stars into corotation and evolution proceeds on a tidal
timescale (i.e., $\tau_{\rm tid}=\frac{\Omega_{\rm spin}}{\dot \Omega_{\rm spin}}$, where the dot indicates a time derivative, $\Omega_{\rm spin}$ is the spin angular velocity, Bagot 1996). As a result, angular momentum
loss would ultimately lead to a lack of corotation and thence to a failure of
the Roche model. Meanwhile, magnetic braking is always accompanied by the
mass loss from the system, and the mass loss has a significant
influence on the evolution of W UMa systems. We will
consider the spin angular momentum and mass loss of the system to
construct a true non-conservative model for low-mass W UMa systems in our
future work.

\section*{ACKNOWLEDGEMENTS}
We acknowledge the generous support provided by the Chinese National Science
Foundation (Grant No. 10273020, 10303006 and 19925312), the Foundation
of Chinese Academy of Sciences (KJCX2-SW-T06),
Yunnan Natural Science Foundation (Grant No. 2002A0020Q), and by 973
Scheme (NKBRSF G19990754).  We are grateful to Dr. C. Tout, the referee,
for his valuable suggestions and insightful remarks which improve this paper greatly and for
his kind help in improving authors' English language.

\bsp

\label{lastpage}

\end{document}